\documentclass[%
 reprint,
 amsmath,amssymb,
 aps,
prb,
]{revtex4-2}

\usepackage{graphicx}
\usepackage{dcolumn}
\usepackage{bm}
\usepackage[colorlinks=true, linkcolor=blue, citecolor=blue, urlcolor=blue]{hyperref}
\usepackage{xcolor}

\begin{document}

\preprint{APS/123-QED}

\title{Predictive Indicator of Critical Point in Equilibrium and Nonequilibrium Magnetic Systems}
\author{Tianyi Zhang$^{1}$, Caihua Wan$^{1,\ 2*}$ and Xiufeng Han$^{1,2,3}$}

 \email{xfhan@iphy.ac.cn; wancaihua@iphy.ac.cn}
\affiliation{%
 ${}^{1}$Beijing National Laboratory for Condensed Matter Physics, University of Chinese Academy of Sciences, Chinese Academy of Sciences, Beijing 100190, China\\
 ${}^{2}$Center of Materials Science and Optoelectronics Engineering, University of Chinese Academy of Sciences, Beijing 100049, China\\
 ${}^{3}$Songshan Lake Materials Laboratory, Dongguan, Guangdong 523808, China
}%

\begin{abstract}
Determining critical points of phase transitions from partial data is essential to avoid abrupt system collapses and reducing experimental or computational costs. However, the complex physical systems and phase transition phenomena have long hindered the development of unified approaches applicable to both equilibrium and nonequilibrium phase transitions. In this work, we propose predictive indicators to determine critical points in equilibrium and nonequilibrium magnetic systems based on frequency-dependent response function. For equilibrium phase transition, such as magnetization switching under magnetic field, the static magnetization response function to a perturbing magnetic field diverges at the critical field, serving as a noise-resilient predictive indicator that also reflects the transition order and critical exponents. In contrast, for nonequilibrium phase transition, such as magnetization switching driven by spin-transfer torque, static response fails to signal criticality. Instead, the dynamic response at ferromagnetic resonance frequency diverges at the critical point, which is also robust against thermal noise. We further demonstrate that these static and dynamic indicators can be unified in the frame of first-order linear differential systems, offering a generalizable strategy for predicting criticality in both equilibrium and nonequilibrium transitions.
\end{abstract}

\maketitle

Phase transition is an emergent \cite{Anderson_1972,Strogatz_2022,Rupe_2024} phenomenon that widely exists in physical systems \cite{Kosterlitz_1973,Kosterlitz_1974,Amit_1985,Zivieri_2016}, biological systems \cite{Dakos_2008,Hughes_2018,Berdugo_2020,Flores_2024,Yang_2025}, and socio-economic networks\cite{Moran_2025}. Predicting the critical point of phase transitions helps avoid sudden system collapses, reduces the cost of experiments and simulations, and aids in understanding emergent behaviors. For example, in spintronics, predicting the critical current \cite{Sun_2000,Lee_2013,Taniguchi_2015,Taniguchi_2016,Zhu_2020,Zhang_2024_prb} is essential for stabilizing magnetic random-access memory (MRAM) devices. Several works have proposed significant methods for predicting critical points, such as mean-field theory \cite{Negele_1982}, Landau's theory of phase transitions \cite{Aranson_2002}, renormalization group \cite{Nienhuis_1975}, machine learning \cite{Wang_2016,Huang_2024,Panahi_2024}, dynamical equation analysis \cite{Sun_2000,Lee_2013,Taniguchi_2015,Taniguchi_2016,Zhu_2020,Zhang_2024_prb}, and critical slowing down and variance increase analysis \cite{Dakos_2008,Scheffer_2009,Scheffer_2012}. These methods require prior knowledge of the system's Hamiltonian or dynamical equations, or necessitate dense sampling near the critical point. For unknown systems, determining critical point based on partial data is an interesting and challenging task.

There are distinctions in the phase transition mechanisms, critical behaviors, and research methods between equilibrium and nonequilibrium phase transitions, while studies on their connections are relatively scarce. Complex research systems and phase transition phenomena have hindered the development of general methods for determining the critical points of both equilibrium and nonequilibrium phase transitions. Magnetic systems, however, exhibit both types of phase transitions and have clear physical pictures. On the one hand, the magnetization can be switched by magnetic field, and the phase transition occurs due to changes in the minimum of the free energy, which belongs to equilibrium phase transitions. On the other hand, in cases where magnetization is switched using spin-transfer torque (STT) \cite{Ralph_2008}, spin-orbit torque (SOT) \cite{Song_2021,Han_2021}, or laser \cite{Kirilyuk_2013,Cheng_2017,Deb_2018,Peng_2023,Lin_2024}, the phase transition arises from dynamic instability caused by the injection of energy flow, which belongs to nonequilibrium phase transitions. Therefore, magnetic systems serve as an ideal platform for studying both types of phase transitions. However, a unified method for predicting the critical points of equilibrium and nonequilibrium phase transitions in magnetic systems is currently lacking and remains an open question.

In this study, we propose two appropriate prediction indicators for critical points in equilibrium and nonequilibrium magnetic systems. We calculate the cases of magnetic field and STT driven ferromagnet switching with perpendicular magnetic anisotropy, corresponding to equilibrium and nonequilibrium phase transitions, and study the response function of magnetization to a transverse disturbance field. Additionally, we explored how to utilize the static response to determine the transition order and critical exponents of equilibrium phase transitions. Finally, we general these two indicators to first-order linear dynamical systems.

\begin{figure}[htbp]
\includegraphics[width = \linewidth]{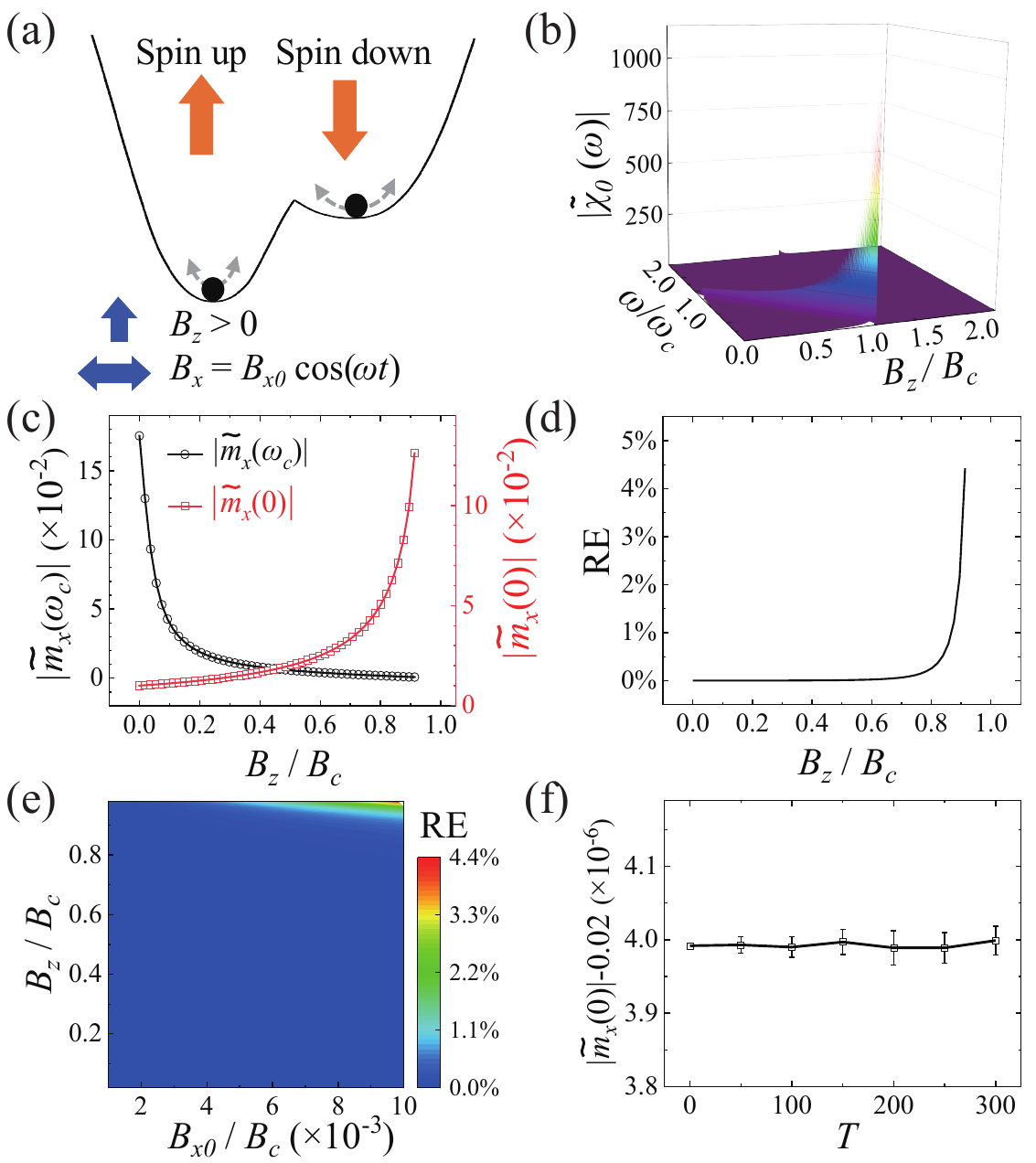}
\caption{\label{fig1}(a) Energy landscape of a double-well potential representing the equilibrium magnetic system under $B_x = B_{x0} \mathrm{cos}(\omega t)$ and $B_z > 0$. (b) The dependence of response function $|\widetilde{\chi}_{0}(\omega)|$ on the dc magnetic field $B_z$ and ac perturbation magnetic field’s frequency $\omega$. (c) The dependence of $|\widetilde{m}_{x}(\omega)|$ on $B_z$ at $\omega$ = 0 and $\omega_c$ for the lower half branch of the hysteresis loop with $B_{x0} = 0.01B_c$. (d) Relative error (RE) between $|\widetilde{m}_{x}(0)|$ and $|\widetilde{m}_{x0}(0)|$. (e) The RE as a function of $B_z$ and $B_{x0}$. (f) The $|\widetilde{m}_{x}(0)|$ dependence on the temperature $T$ at $B_z = 0.5B_c$.}
\end{figure}

Predicting the critical magnetic field for magnetization switching is crucial for designing new magnetic materials and improving magnetic device stability. For a ferromagnet with perpendicular magnetic anisotropy subjected to a constant magnetic field $B_z$ in the z-direction and an oscillatory perturbative field $B_x = B_{x0} \mathrm{cos}(\omega t)$ in the x-direction, the double-well potential model describing the system's energy is illustrated in Fig. \ref{fig1}a. If the initial state is spin down, as the $B_z$ gradually increases, the energy of the spin-down state gradually increases. When it exceeds the barrier, the magnetization undergoes a switching, which is a equilibrium phase transition process, and the state of the system is determined by minimizing the free energy. In order to study the effect of Bx on the magnetization, we calculated the response function of $m_x$ to $B_x$, which is defined as
\begin{equation}
\chi_{0}\left(t-t^{\prime}\right)=\frac{m_{x}(t)}{B_{x}\left(t^{\prime}\right)}
\end{equation}
Its Fourier transform $\widetilde{\chi}_{0}(\omega)$ reflects the amplitude of the $m_x$ oscillation at frequency $\omega$. By first linearizing the Landau-Lifshitz-Gilbert (LLG) equation up to the first-order terms of $m_x$ and $m_y$, then performing the Fourier transformation on the equation, and finally calculating the frequency response of $m_x$ with respect to $B_x$, one can obtain
\begin{equation}
\widetilde{\chi}_{0}(\omega)=\frac{\alpha^{2}\left(B_{z} m_{z}+K m_{z}^{2}\right)+B_{z}m_{z}+K -i \frac{\mu_{S} \alpha\left(1+\alpha^{2}\right)}{\gamma} \omega}{\left[\alpha\left(B_{z} m_{z}+K m_{z}^{2}\right)-i \frac{\mu_{s}\left(1+\alpha^{2}\right])}{\gamma} \omega\right]^{2}+\left(B_{z}+K m_{z}\right)^{2}}
\label{eq2}  
\end{equation}
Here $K$ is anisotropy field, $\mu_s$ is saturation magnetization and $\gamma$ is gyromagnetic ratio. For a detailed derivation, please refer to Supplemental material A \cite{sup2025}. The dependence of $|\widetilde{\chi}_{0}(\omega)|$ on $\omega$ and $B_z$ is shown in Fig. \ref{fig1}(b). Here, $B_c = K$ represents the critical magnetic field, and $\omega_{c}=\gamma K / \mu_{s}$ represents the ferromagnetic resonance frequency in the absence of an external magnetic field, $B_{x0} = 0.01B_c$. As shown in Fig. \ref{fig1}(b), $|\widetilde{\chi}_{0}(0)|$ increases rapidly as $B_z$ approaches $B_c$, which suggests that we can use $|\widetilde{m}_{x}(0)|$ as a predictive indicator for the magnetization switching. Substituting  $\omega = 0$ and $m_z=-1$ into Eq. (\ref{eq2}), we can obtain
\begin{equation}
\widetilde{\chi}_{0}(0)=\frac{1}{-B_{z}+K}
\end{equation}
The relationship between $|\widetilde{\chi}_{0}(0)|$ and $|\widetilde{\chi}_{0}(\omega_c)|$ with respect to $B_z$ for the lower half branch of the hysteresis loop is shown in Fig. \ref{fig1}(c). We can observe that $|\widetilde{\chi}_{0}(0)|$ increases rapidly as $B_z$ approaches $B_c$, while $|\widetilde{\chi}_{0}(\omega_c)|$ gradually decays, which is consistent with the results of past experiments \cite{Mathews_2021,Jungfleisch_2015} and theoretical studies \cite{Chantrell_1989,Cookson_2002,Moorfield_2020}. This is because the increase in $B_z$ destroys the resonance condition. The relative error RE = $\left|\widetilde{\chi}(0)-\widetilde{\chi}_{0}(0)\right| /\left(\widetilde{\chi}(0)+\widetilde{\chi}_{0}(0)\right)$, where $\widetilde{\chi}(0)$ is simulated response function using LLG equation, with respect to $B_z$ varies as shown in Fig. \ref{fig1}(d). Near $B_c$, RE increases rapidly, because the deviation of magnetization becomes larger near $B_c$, violating the assumption $|m_x| \ll 1$. The dependence of RE on $B_z$ and $B_{x0}$ is shown in Fig. \ref{fig1}(e). We can observe that as $B_z$ approaches $B_c$ and $B_{x0}$ gradually increases, the RE increases, which is also due to the increase magnetization deviation. Finally, thermal noise is introduced and the temperature $T$ dependence of $|\widetilde{m}_{x}(0)|$ is calculated at $B_z = 0.5B_c$, as shown in Fig. \ref{fig1}(f). Each temperature is repeated 20 times. It shows the temperature has almost no effect on the mean value of $|\widetilde{m}_{x}(0)|$, but the variance increases as the temperature rises. This demonstrates the temperature robustness of $|\widetilde{m}_{x}(0)|$ as a predictive indicator for magnetization switching. 

\begin{figure}[htbp]
\includegraphics[width = \linewidth]{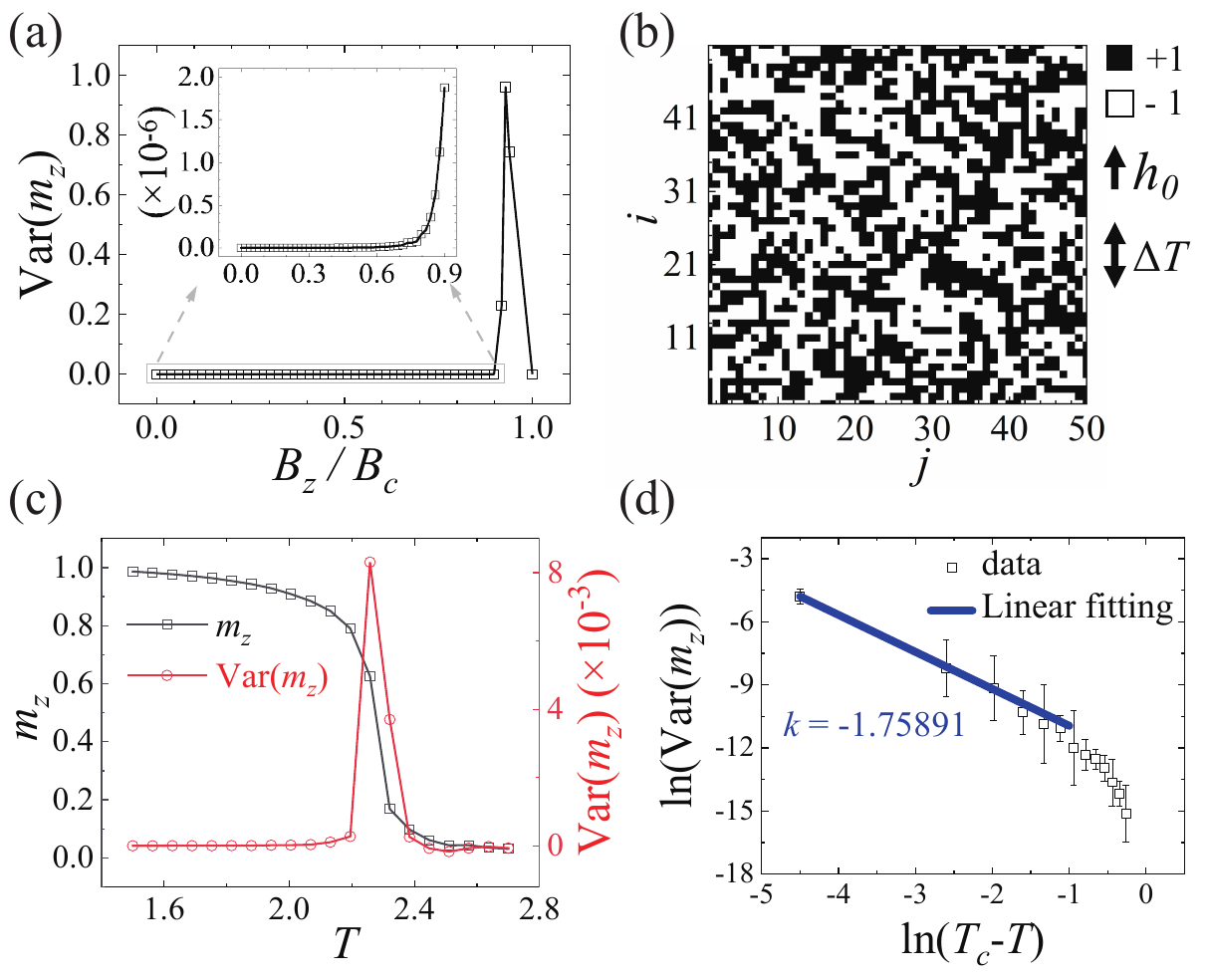}
\caption{\label{fig2}(a) The perturbation-induced variance Var($m_z$) of the lower half part of hysteresis loop as a function of $B_z$. (b) A 50$\times$50 Ising model subject to a constant magnetic field $h_0$ and a perturbation temperature $\Delta T$. (c) A combined plot of magnetization $m_z$ and Var($m_z$) as a function of $T$ with $h_0$ = 0 and $\Delta T_0 = 0.2$. (d) The relationship between Var($m_z$) and temperature $T$ near the critical point $T_c \approx 2.269$.}
\end{figure}

In the previous section, it is demonstrated that the response of an equilibrium magnetic system to a static perturbation magnetic field is amplified near the critical point. Interestingly, by utilizing this property, we can determine the transition order and the critical exponent of the equilibrium phase transition. For the first-order phase transition system shown in Fig. \ref{fig1}(a), by linearizing the equilibrium LLG equation, solves analytically for $m_z$ under a small uniform transverse perturbation $B_x$ from $-B_{x0}$ to $B_{x0}$, and computes the variances of $m_z$ as
\begin{equation}
\operatorname{Var}\left(m_{z}\right)=2 B_{x 0}^{4} /\left[45\left(B_{z}-B_{c}\right)^{4}\right]
\end{equation}
And the max value is
\begin{equation}
\operatorname{Var}\left(m_{z}\right)_{max}=1
\end{equation}
For a detailed derivation, please refer to Supplemental material B \cite{sup2025}. The dependence of Var($m_z$) on $B_z$ is shown in Fig. \ref{fig2}(a). Worth noting, the maximum value of Var($m_z$) does not depend on $B_{x0}$, but only on the difference in magnetization before and after the critical point. For the case of a second-order phase transition, we take the ferromagnetic-to-paramagnetic transition in two-dimensional Ising model as an example. As shown in Fig. \ref{fig2}(b), a 50×50 Ising model is subject to a constant magnetic field $h_0$ and a perturbation temperature $\Delta T$, which follows a uniform distribution from $ -\Delta T_0$ to $ \Delta T_0$. The magnetization $m_z$ and $\operatorname{Var}\left(m_{z}\right)$ as a function of temperature are shown in Fig. \ref{fig2}(c) with exchange interaction energy $J = 1$, Boltzmann's constant $k_B =1$, $h_0$ = 0 and $\Delta T_0$ = 0.2. Near the critical temperature $T_c$ = 2.269, the magnetization approximately follows $m_{z} \propto\left(T_{c}-T\right)^{\alpha}$, where $\alpha$ is the critical exponent ideally equals to 0.125 \cite{Yang_1952}. The dependence of Var($m_z$) on $T$ is:
\begin{equation}
\operatorname{Var}\left(m_{z}\right)=\frac{1}{3}\left(T_{c}-T\right)^{2 \alpha-2} \Delta T_{0}^{2}
\end{equation}
And the max value
\begin{equation}
\operatorname{Var}\left(m_{z}\right)_{\max } \propto\left(\Delta T_{0}\right)^{2 \alpha}
\end{equation}
By comparing Eqs. (4) and (6), Eqs. (5) and (7), two interesting conclusions can be extracted. Firstly, while the peak value of the variance remains independent of perturbation strength in first-order transitions, it strongly depends on it in second-order cases. Secondly, in the case of second-order transitions, the critical exponent $\alpha$ can be extracted by examining the scaling behavior of the variance near the critical point. Specifically, the scaling behavior of Var($m_z$) is plotted on a logarithmic scale and performed a linear fit in the vicinity of the transition critical point, as shown in Fig. \ref{fig2}(d). The estimate $\alpha \approx $0.121, which is in close agreement with the theoretical prediction 0.125. Deviations from linearity observed far from the critical point arise due to the breakdown of the approximation $m_{z}=\left[1-\sinh ^{-4}\left(\ln (1+\sqrt{2}) T_{c} / T\right)\right]^{1 / 8} \approx\left(1-T / T_{c}\right)^{1 / 8}$, which holds only in the regime $(T_c-T)/T_c \ll 1$.

\begin{figure}[htbp]
\includegraphics[width = \linewidth]{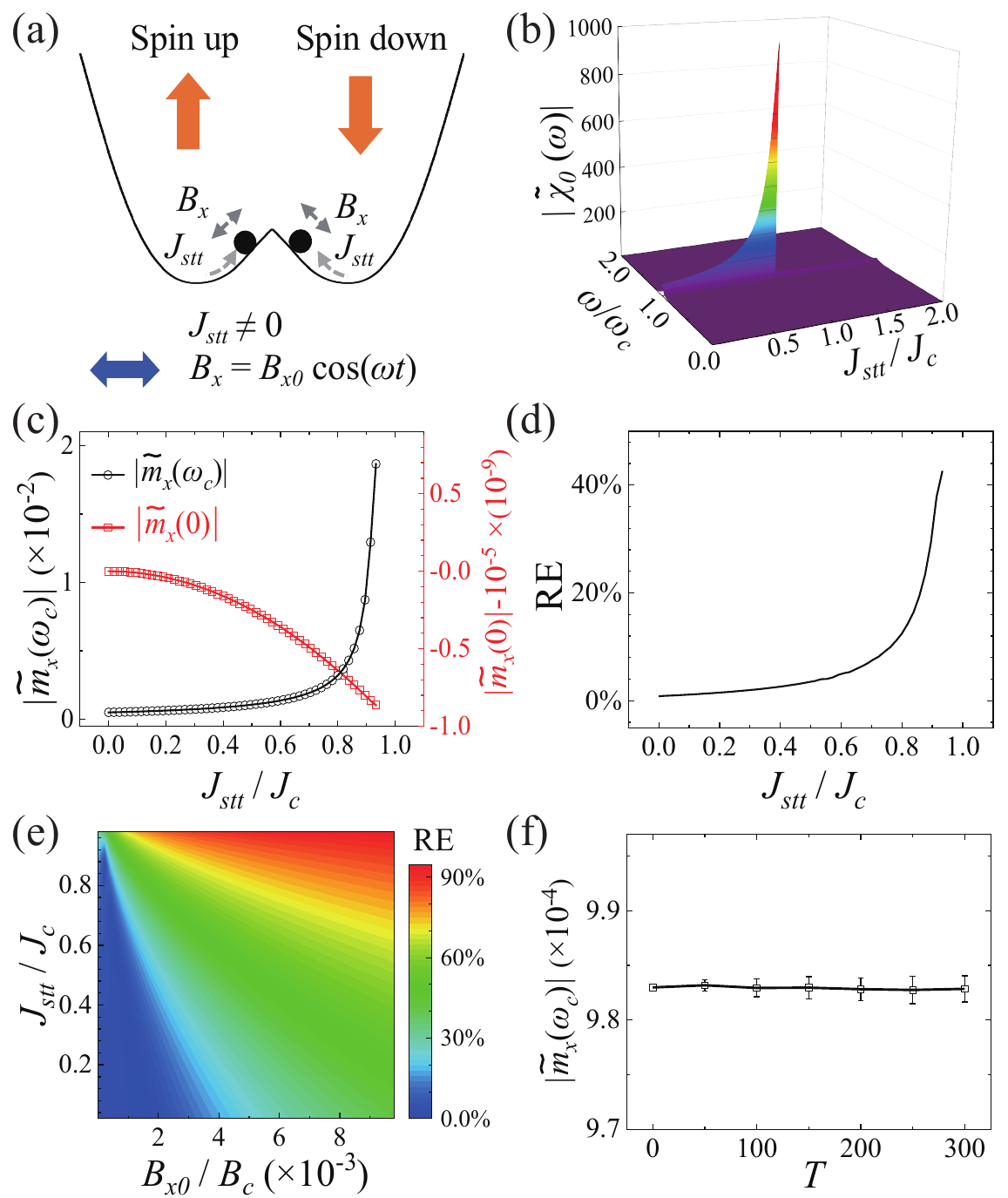}
\caption{\label{fig3}(a) Energy landscape of a double-well potential representing the nonequilibrium magnetic system under $B_x = B_{x0} \mathrm{cos}(\omega t)$ and spin transfer torque strength $J_{stt}$. (b) The dependence of response function $|\widetilde{\chi}_{0}(\omega)|$ on the $J_{stt}$ and ac perturbation magnetic field’s frequency $\omega$. (c) The dependence of $|\widetilde{m}_{x}(\omega)|$ on $J_{stt}$ at $\omega$ = 0 and $\omega_c$ for the lower half branch of the hysteresis loop with $B_{x0} = 10^{-5}B_c$. (d) Relative error (RE) between $|\widetilde{m}_{x}(\omega_c)|$ and $|\widetilde{m}_{x0}(\omega_c)|$. (e) The RE as a function of $J_{stt}$ and $B_{x0}$. (f) The $|\widetilde{m}_{x}(\omega_c)|$ dependence on the temperature $T$ at $J_{stt} = 0.5J_c$.}
\end{figure}

Compared to equilibrium phase transitions, due to the lack of a unified theoretical framework and complexity of research systems and phenomena, nonequilibrium phase transitions have been relatively less studied. However, ferromagnetic systems can serve as a simple and clear physical model. Consider a ferromagnet with perpendicular magnetic anisotropy subjected to a constant STT $J_{stt}$ polarized in the z-direction, a constant magnetic field $B_z$ in the z-direction and an oscillatory perturbation magnetic field $B_x = B_{x0} \mathrm{cos}(\omega t)$ along the x-direction, the double-well potential model is illustrated in Fig. \ref{fig3}(a). Unlike magnetic field-driven magnetization switching, STT-driven magnetization switching occurs by destabilizing the system from its equilibrium point rather than altering the system's energy landscape. When the STT is sufficiently large to push the system over the energy barrier, the magnetization switches, which belongs to nonequilibrium phase transition process. And conclusions drawn from previous studies on equilibrium phase transitions valid. Therefore, by using the same method as in Eq. (\ref{eq2}), and taking into account the influence of STT in the LLG equation, we can obtain the response function as
\begin{widetext}
\begin{equation}
\widetilde{\chi}_{0}(\omega)=\frac{\alpha\left(J_{s t t} m_{z}+\alpha\left(B_{z} m_{z}+K m_{z}^{2}\right)\right)+B_{z}m_z-\alpha J_{s t t}m_z+K-i \frac{\mu_{s}\alpha \left(1+\alpha^{2}\right)}{\gamma} \omega}{\left(J_{s t t} m_{z}+\alpha\left(B_{z} m_{z}+K m_{z}^{2}\right)-i \frac{\mu_{s}\left(1+\alpha^{2}\right)}{\gamma} \omega\right)^{2}+\left(B_{z}-\alpha J_{s t t}+K m_{z}\right)^{2}}
\label{eq8}  
\end{equation}
\end{widetext}
Substituting  $\omega = 0$ and $m_z=-1$ into Eq. (\ref{eq8}), we can obtain
\begin{equation}
\widetilde{\chi}_{0}(0)=\frac{K-B_{z}}{J_{s t t}^{2}+\left(B_{z}-K\right)^{2}}
\label{eq9}  
\end{equation}
Detailed derivation is in Supplemental material C \cite{sup2025}. From Eq. (\ref{eq9}), it can be seen that as the absolute value of $J_{stt}$ increases, the static response function decreases monotonically. Clearly, it cannot serve as an indicator for predicting phase transitions. Therefore, we analyze the situation at the critical current, substituting $J_{stt} = J_c = -\alpha(B_z + K m_z)$ and $m_z = -1$ in Eq. (\ref{eq8}). To simplify the physical picture, we consider the case where $B_z =0$. It follows that
\begin{equation}
\left.\widetilde{\chi}_{0}(\omega)\right|_{J_{s t t}=J_{c}}=\frac{K+i \alpha \frac{\mu_{s}}{\gamma} \omega}{\left(\alpha^{2}+1\right)\left(K^{2}-\left(\frac{\mu_{s}}{\gamma} \omega\right)^{2}\right)}
\end{equation}
Therefore, when $\omega = \omega_c$, the response function tends to diverge at the critical point. And $|\widetilde{\chi}_{0}(\omega_c)|$ is
\begin{equation}
|\tilde{\chi}_{0}\left(\omega_{c}\right)|=\frac{\sqrt{1+\alpha^{2} }K}{\left(\alpha K-J_{s t t}\right) \sqrt{\left(J_{s t t}+\alpha K\right)^{2}+4 K^{2}}}
\end{equation}
The dependence of the response function on $\omega$ and $J_{stt}$ is shown in Fig. \ref{fig3}(b). It can be observed that $|\widetilde{\chi}_{0}(\omega_c)|$ increases rapidly as $J_{stt}$ approaches $J_c$, indicating that we can use $|\widetilde{m}_{x}(\omega_c)|$ as a predictive indicator for STT-driven magnetization switching. The dependence of $|\widetilde{m}_{x}(0)|$ and $|\widetilde{m}_{x}(\omega_c)|$ on $J_{stt}$ for the lower half of the hysteresis loop is shown in Fig. \ref{fig3}(c). As $J_{stt}$ approaches $J_c$, $|\widetilde{m}_{x}(\omega_c)|$ increases rapidly, while  $|\widetilde{m}_{x}(0)|$gradually decays, which is exactly the opposite of the situation where the magnetization switched by magnetic field. It is consistent with the results of past experiments \cite{Wang_2018,Petit_2007}. Interestingly, it shows that when the magnetic system is in a resonant state, its response to external disturbances is amplified, while its response to internal disturbances is suppressed. We calculated the relative error RE, as shown in Fig. \ref{fig3}(d). Near $J_c$, RE increases rapidly because the magnetization deviation from the equilibrium state increases. It is worth noting that for the same $B_{x0}$, the magnetization in the resonant state deviates from the equilibrium state much more than not in the resonant state. Therefore, in the calculations of Fig. \ref{fig3}(c)–(f), $B_{x0} = 10^{-5}B_c$, which is 1/1000 of that in Fig. \ref{fig1}. The dependence of RE on $J_{stt}$ and $B_{x0}$ is shown in Fig. \ref{fig3}(e). It can be seen that as $J_{stt}$ and $B_{x0}$ increase, RE increases rapidly, which is also because of the increased magnetization deviation. Finally, we calculated the effect of thermal noise on $|\widetilde{m}_{x}(\omega_c)|$ at $J_{stt}$ = 0.5$J_c$, as shown in Fig. \ref{fig3}(f). By repeating the calculation 20 times at each temperature, it can be observed that as the temperature increases, the mean value of $|\widetilde{m}_{x}(\omega_c)|$ remains essentially unchanged, but the variance gradually increases, which is similar to the situation when the magnetic field switched the magnetization, reflecting the temperature robustness of $|\widetilde{m}_{x}(\omega_c)|$ as a critical prediction indicator. Dynamic response under weak transverse perturbation provides as a sensitive indicator of approaching the critical point in MRAM devices, offering an early warning of write/read failure, instability and inhomogeneity. This method also establishes a predictive and stabilizing mechanism for reliable spin torque oscillator operation.

Based on the above analysis, it is demonstrated that for magnetization switching driven by magnetic fields and STT, the response amplitude of $m_x$ under perturbation fields at frequencies of 0 and $\omega_c$ can be served as predictive indicators for critical points. It is interesting to generalize this approach to more general systems. For a system exhibiting both equilibrium and nonequilibrium phase transitions, assume it follows a first-order linear differential equation,
\begin{equation}
\frac{d \mathbf{X}(t)}{d t}=A \mathbf{X}(t)+\mathbf{B}(t)
\end{equation}
In the equation, $\mathbf{X}$ is an $n$-dimensional vector composed of order parameters, $A$ is the dynamic coefficient matrix, $\mathbf{B}$ is input vector that determines the steady state of the system. The eigenvalues $\lambda_i ( i = 1, 2,\cdots, n)$ of $A$ are ordered according to their real parts, with the assumption that $\operatorname{Re}\left(\lambda_{1}\right) \geq \operatorname{Re}\left(\lambda_{2}\right) \geq \cdots \geq \operatorname{Re}\left(\lambda_{n}\right)$. When the system is at the critical point of an equilibrium phase transition, it must hold that $\operatorname{Re}\left(\lambda_{1}\right)=0$. Since the system is in equilibrium and there is no external energy input, oscillations cannot occur at the critical state. Therefore, $\operatorname{Im}\left(\lambda_{1}\right)=0$ holds true and det($A$) = 0, which implies that for a static small perturbation, the response of $\mathbf{X}$, which is $A^{-1}\mathbf{B}$, tends to diverge as system approaches the critical point. For example, in the case of magnetic field-driven magnetization switching, $\lambda_{1,2}=-\gamma /\left(\mu_{s}\left(1+\alpha^{2}\right)\right)\left[\alpha\left(-B_{z}+K\right) \pm \mathrm{i}\left(B_{z}-K\right)\right]$, $\quad \mathbf{B}(\mathrm{t})=\gamma /\left(\mu_{s}\left(1+\alpha^{2}\right)\right)(\alpha, 1)^{\mathrm{T}} B_{x 0}$, at the critical magnetic field $B_c=K$, $\operatorname{Re}\left(\lambda_{1}\right)=\operatorname{Im}\left(\lambda_{1}\right)=0$. However, for nonequilibrium phase transitions, the situation is totally different. At the critical point, there still exists $\operatorname{Re}\left(\lambda_{1}\right)=0$, but injection of external energy flow enables the oscillatory behavior, which is a main difference between equilibrium and nonequilibrium phase transitions. Therefore, $\operatorname{Im}\left(\lambda_{1}\right)$ may not equals to 0 at critical point and Hopf bifurcation may occur. For a constant perturbation, the response of $\mathbf{X}$ may not tend to diverge at the critical point. For instance, in the case of STT-driven magnetization switching discussed above, $\lambda_{1,2}=-\gamma /\left(\mu_{s}\left(1+\alpha^{2}\right)\right)\left[-J_{S T T}+\alpha\left(K-B_{z}\right) \pm \mathrm{i}\left(B_{z}-\alpha J_{S T T}\right.\right.$ $-K)]$, and at the critical current $J_{c}=\alpha\left(K-B_{z}\right)$, $\operatorname{Re}\left(\lambda_{1,2}\right)=0$, $\operatorname{Im}\left(\lambda_{1,2}\right)= \pm \mathrm{i} \gamma\left(K-B_{z}\right) / \mu_{s}$. However, if the perturbation has a frequency of $\omega_c =| \operatorname{Im}\left(\lambda_{1}\right)|$, the response of $\mathbf{X}$ will gradually diverge as system approaches the critical point. Therefore, $\widetilde{\mathbf{X}}(0)$ and $\widetilde{\mathbf{X}}(\omega_c)$ can serve as predictive indicators for critical point in equilibrium and nonequilibrium phase transition, respectively. Several significant works previously reported are also consistent with this conclusion. For instance, the static response function in the desynchronized-synchronized equilibrium transition of the Kuramoto model \cite{Roberts_2008,Garc_2017} diverges at the critical point.
It is worth noting that in the quantum system (for example, Rydberg atoms\cite{Zhang_2024}), we also observed the divergent behavior of dynamic indicators near the critical point, indicating the potential application of our theory in quantum physics. However, in topological phase transitions without local order parameters, the proposed method in this work is not applicable. And how to predict critical points in these systems is worthy of further investigation.

In summary, we have proposed two prediction indicators for the critical points of magnetization switching driven by magnetic fields and STT in magnetic systems. In the case of magnetic field-driven switching, by applying a transverse static disturbance field $B_{x0}$, the response $m_x$ diverges as $B_z$ approaches the critical point $B_c$, serving as a noise-resilient predictive indicator. By replacing the static disturbance with a uniformly distributed disturbance and performing multiple measurements to obtain the variance of the order parameter, we found that for first-order phase transitions, the maximum variance depends only on the difference of the order parameter across the critical point and is independent of the disturbance magnitude. However, for second-order phase transitions, the maximum variance is proportional to the 2$\alpha$-th power of the disturbance magnitude. Near the critical point, the variance is proportional to the 2($\alpha$-1)-th power of the difference between the critical point and the variable. Using this property, we calculated the temperature critical exponent of a 50 × 50 two-dimensional Ising model and get 0.121, which is close to the theoretical value of 0.125. For STT-driven magnetization switching, applying a transverse disturbance field oscillating at the ferromagnetic resonance frequency, the response $m_x$ at frequency $\omega_c$ rapidly increases as $J_{stt}$ approaches the critical point $J_c$, also serving as a noise-resilient predictive indicator. Finally, we demonstrate that this framework is applicable to general first-order linear systems. Our approach offers predictive indicators for determining critical points in both equilibrium and nonequilibrium phase transitions, which is computationally and experimentally cost-effective, easy to implement, and robust against thermal fluctuations. Our work has potential applications in enhanced magnetic sensor, MRAM device test, dynamic control of spin torque oscillators, designing new magnetic materials and serving as warning signals of critical point in complex systems.

\begin{acknowledgments}
This work is financially supported by the National Key Research and Development Program of China (MOST) (Grant No. 2022YFA1402800), the National Natural Science Foundation of China (NSFC) (Grants No. T2495210, No. 51831012, No. 12134017, No. 12374131, and No. W2412079 ) and the Chinese Academy of Sciences President’s International Fellowship Initiative (PIFI Grant No. 2025PG0006).
\end{acknowledgments}

\providecommand{\noopsort}[1]{}\providecommand{\singleletter}[1]{#1}%


\begin{thebibliography}{47}%
\makeatletter
\providecommand \@ifxundefined [1]{%
 \@ifx{#1\undefined}
}%
\providecommand \@ifnum [1]{%
 \ifnum #1\expandafter \@firstoftwo
 \else \expandafter \@secondoftwo
 \fi
}%
\providecommand \@ifx [1]{%
 \ifx #1\expandafter \@firstoftwo
 \else \expandafter \@secondoftwo
 \fi
}%
\providecommand \natexlab [1]{#1}%
\providecommand \enquote  [1]{``#1''}%
\providecommand \bibnamefont  [1]{#1}%
\providecommand \bibfnamefont [1]{#1}%
\providecommand \citenamefont [1]{#1}%
\providecommand \href@noop [0]{\@secondoftwo}%
\providecommand \href [0]{\begingroup \@sanitize@url \@href}%
\providecommand \@href[1]{\@@startlink{#1}\@@href}%
\providecommand \@@href[1]{\endgroup#1\@@endlink}%
\providecommand \@sanitize@url [0]{\catcode `\\12\catcode `\$12\catcode `\&12\catcode `\#12\catcode `\^12\catcode `\_12\catcode `\%12\relax}%
\providecommand \@@startlink[1]{}%
\providecommand \@@endlink[0]{}%
\providecommand \url  [0]{\begingroup\@sanitize@url \@url }%
\providecommand \@url [1]{\endgroup\@href {#1}{\urlprefix }}%
\providecommand \urlprefix  [0]{URL }%
\providecommand \Eprint [0]{\href }%
\providecommand \doibase [0]{https://doi.org/}%
\providecommand \selectlanguage [0]{\@gobble}%
\providecommand \bibinfo  [0]{\@secondoftwo}%
\providecommand \bibfield  [0]{\@secondoftwo}%
\providecommand \translation [1]{[#1]}%
\providecommand \BibitemOpen [0]{}%
\providecommand \bibitemStop [0]{}%
\providecommand \bibitemNoStop [0]{.\EOS\space}%
\providecommand \EOS [0]{\spacefactor3000\relax}%
\providecommand \BibitemShut  [1]{\csname bibitem#1\endcsname}%
\let\auto@bib@innerbib\@empty
\bibitem [{\citenamefont {Anderson}(1972)}]{Anderson_1972}%
  \BibitemOpen
  \bibfield  {author} {\bibinfo {author} {\bibfnamefont {P.~W.}\ \bibnamefont {Anderson}},\ }\bibfield  {title} {\bibinfo {title} {More is different},\ }\href {https://doi.org/10.1126/science.177.4047.393} {\bibfield  {journal} {\bibinfo  {journal} {Science}\ }\textbf {\bibinfo {volume} {177}},\ \bibinfo {pages} {393} (\bibinfo {year} {1972})}\BibitemShut {NoStop}%
\bibitem [{\citenamefont {Strogatz}\ \emph {et~al.}(2022)\citenamefont {Strogatz}, \citenamefont {Walker}, \citenamefont {Yeomans}, \citenamefont {Tarnita}, \citenamefont {Arcaute}, \citenamefont {De~Domenico}, \citenamefont {Artime},\ and\ \citenamefont {Goh}}]{Strogatz_2022}%
  \BibitemOpen
  \bibfield  {author} {\bibinfo {author} {\bibfnamefont {S.}~\bibnamefont {Strogatz}}, \bibinfo {author} {\bibfnamefont {S.}~\bibnamefont {Walker}}, \bibinfo {author} {\bibfnamefont {J.~M.}\ \bibnamefont {Yeomans}}, \bibinfo {author} {\bibfnamefont {C.}~\bibnamefont {Tarnita}}, \bibinfo {author} {\bibfnamefont {E.}~\bibnamefont {Arcaute}}, \bibinfo {author} {\bibfnamefont {M.}~\bibnamefont {De~Domenico}}, \bibinfo {author} {\bibfnamefont {O.}~\bibnamefont {Artime}},\ and\ \bibinfo {author} {\bibfnamefont {K.-I.}\ \bibnamefont {Goh}},\ }\bibfield  {title} {\bibinfo {title} {Fifty years of ‘more is different’},\ }\href {https://doi.org/10.1038/s42254-022-00483-x} {\bibfield  {journal} {\bibinfo  {journal} {Nature Reviews Physics}\ }\textbf {\bibinfo {volume} {4}},\ \bibinfo {pages} {508} (\bibinfo {year} {2022})}\BibitemShut {NoStop}%
\bibitem [{\citenamefont {Rupe}\ and\ \citenamefont {Crutchfield}(2024)}]{Rupe_2024}%
  \BibitemOpen
  \bibfield  {author} {\bibinfo {author} {\bibfnamefont {A.}~\bibnamefont {Rupe}}\ and\ \bibinfo {author} {\bibfnamefont {J.~P.}\ \bibnamefont {Crutchfield}},\ }\bibfield  {title} {\bibinfo {title} {On principles of emergent organization},\ }\href {https://doi.org/https://doi.org/10.1016/j.physrep.2024.04.001} {\bibfield  {journal} {\bibinfo  {journal} {Physics Reports}\ }\textbf {\bibinfo {volume} {1071}},\ \bibinfo {pages} {1} (\bibinfo {year} {2024})}\BibitemShut {NoStop}%
\bibitem [{\citenamefont {Kosterlitz}\ and\ \citenamefont {Thouless}(1973)}]{Kosterlitz_1973}%
  \BibitemOpen
  \bibfield  {author} {\bibinfo {author} {\bibfnamefont {J.~M.}\ \bibnamefont {Kosterlitz}}\ and\ \bibinfo {author} {\bibfnamefont {D.~J.}\ \bibnamefont {Thouless}},\ }\bibfield  {title} {\bibinfo {title} {Ordering, metastability and phase transitions in two-dimensional systems},\ }\href {https://doi.org/10.1088/0022-3719/6/7/010} {\bibfield  {journal} {\bibinfo  {journal} {Journal of Physics C: Solid State Physics}\ }\textbf {\bibinfo {volume} {6}},\ \bibinfo {pages} {1181} (\bibinfo {year} {1973})}\BibitemShut {NoStop}%
\bibitem [{\citenamefont {Kosterlitz}(1974)}]{Kosterlitz_1974}%
  \BibitemOpen
  \bibfield  {author} {\bibinfo {author} {\bibfnamefont {J.~M.}\ \bibnamefont {Kosterlitz}},\ }\bibfield  {title} {\bibinfo {title} {The critical properties of the two-dimensional xy model},\ }\href {https://doi.org/10.1088/0022-3719/7/6/005} {\bibfield  {journal} {\bibinfo  {journal} {Journal of Physics C: Solid State Physics}\ }\textbf {\bibinfo {volume} {7}},\ \bibinfo {pages} {1046} (\bibinfo {year} {1974})}\BibitemShut {NoStop}%
\bibitem [{\citenamefont {Amit}\ \emph {et~al.}(1985)\citenamefont {Amit}, \citenamefont {Gutfreund},\ and\ \citenamefont {Sompolinsky}}]{Amit_1985}%
  \BibitemOpen
  \bibfield  {author} {\bibinfo {author} {\bibfnamefont {D.~J.}\ \bibnamefont {Amit}}, \bibinfo {author} {\bibfnamefont {H.}~\bibnamefont {Gutfreund}},\ and\ \bibinfo {author} {\bibfnamefont {H.}~\bibnamefont {Sompolinsky}},\ }\bibfield  {title} {\bibinfo {title} {Storing infinite numbers of patterns in a spin-glass model of neural networks},\ }\href {https://doi.org/10.1103/PhysRevLett.55.1530} {\bibfield  {journal} {\bibinfo  {journal} {Physical Review Letters}\ }\textbf {\bibinfo {volume} {55}},\ \bibinfo {pages} {1530} (\bibinfo {year} {1985})}\BibitemShut {NoStop}%
\bibitem [{\citenamefont {Zivieri}(2016)}]{Zivieri_2016}%
  \BibitemOpen
  \bibfield  {author} {\bibinfo {author} {\bibfnamefont {R.}~\bibnamefont {Zivieri}},\ }\bibfield  {title} {\bibinfo {title} {Critical phenomena in ferromagnetic antidot lattices},\ }\href@noop {} {\bibfield  {journal} {\bibinfo  {journal} {AIP Advances}\ }\textbf {\bibinfo {volume} {6}} (\bibinfo {year} {2016})}\BibitemShut {NoStop}%
\bibitem [{\citenamefont {Dakos}\ \emph {et~al.}(2008)\citenamefont {Dakos}, \citenamefont {Scheffer}, \citenamefont {van Nes}, \citenamefont {Brovkin}, \citenamefont {Petoukhov},\ and\ \citenamefont {Held}}]{Dakos_2008}%
  \BibitemOpen
  \bibfield  {author} {\bibinfo {author} {\bibfnamefont {V.}~\bibnamefont {Dakos}}, \bibinfo {author} {\bibfnamefont {M.}~\bibnamefont {Scheffer}}, \bibinfo {author} {\bibfnamefont {E.~H.}\ \bibnamefont {van Nes}}, \bibinfo {author} {\bibfnamefont {V.}~\bibnamefont {Brovkin}}, \bibinfo {author} {\bibfnamefont {V.}~\bibnamefont {Petoukhov}},\ and\ \bibinfo {author} {\bibfnamefont {H.}~\bibnamefont {Held}},\ }\bibfield  {title} {\bibinfo {title} {Slowing down as an early warning signal for abrupt climate change},\ }\href {https://doi.org/10.1073/pnas.0802430105} {\bibfield  {journal} {\bibinfo  {journal} {Proceedings of the National Academy of Sciences}\ }\textbf {\bibinfo {volume} {105}},\ \bibinfo {pages} {14308} (\bibinfo {year} {2008})}\BibitemShut {NoStop}%
\bibitem [{\citenamefont {Hughes}\ \emph {et~al.}(2018)\citenamefont {Hughes}, \citenamefont {Kerry}, \citenamefont {Baird}, \citenamefont {Connolly}, \citenamefont {Dietzel}, \citenamefont {Eakin}, \citenamefont {Heron}, \citenamefont {Hoey}, \citenamefont {Hoogenboom}, \citenamefont {Liu}, \citenamefont {McWilliam}, \citenamefont {Pears}, \citenamefont {Pratchett}, \citenamefont {Skirving}, \citenamefont {Stella},\ and\ \citenamefont {Torda}}]{Hughes_2018}%
  \BibitemOpen
  \bibfield  {author} {\bibinfo {author} {\bibfnamefont {T.~P.}\ \bibnamefont {Hughes}}, \bibinfo {author} {\bibfnamefont {J.~T.}\ \bibnamefont {Kerry}}, \bibinfo {author} {\bibfnamefont {A.~H.}\ \bibnamefont {Baird}}, \bibinfo {author} {\bibfnamefont {S.~R.}\ \bibnamefont {Connolly}}, \bibinfo {author} {\bibfnamefont {A.}~\bibnamefont {Dietzel}}, \bibinfo {author} {\bibfnamefont {C.~M.}\ \bibnamefont {Eakin}}, \bibinfo {author} {\bibfnamefont {S.~F.}\ \bibnamefont {Heron}}, \bibinfo {author} {\bibfnamefont {A.~S.}\ \bibnamefont {Hoey}}, \bibinfo {author} {\bibfnamefont {M.~O.}\ \bibnamefont {Hoogenboom}}, \bibinfo {author} {\bibfnamefont {G.}~\bibnamefont {Liu}}, \bibinfo {author} {\bibfnamefont {M.~J.}\ \bibnamefont {McWilliam}}, \bibinfo {author} {\bibfnamefont {R.~J.}\ \bibnamefont {Pears}}, \bibinfo {author} {\bibfnamefont {M.~S.}\ \bibnamefont {Pratchett}}, \bibinfo {author} {\bibfnamefont {W.~J.}\ \bibnamefont {Skirving}}, \bibinfo {author} {\bibfnamefont {J.~S.}\ \bibnamefont {Stella}},\ and\ \bibinfo
  {author} {\bibfnamefont {G.}~\bibnamefont {Torda}},\ }\bibfield  {title} {\bibinfo {title} {Global warming transforms coral reef assemblages},\ }\href {https://doi.org/10.1038/s41586-018-0041-2} {\bibfield  {journal} {\bibinfo  {journal} {Nature}\ }\textbf {\bibinfo {volume} {556}},\ \bibinfo {pages} {492} (\bibinfo {year} {2018})}\BibitemShut {NoStop}%
\bibitem [{\citenamefont {Berdugo}\ \emph {et~al.}(2020)\citenamefont {Berdugo}, \citenamefont {Delgado-Baquerizo}, \citenamefont {Soliveres}, \citenamefont {Hernández-Clemente}, \citenamefont {Zhao}, \citenamefont {Gaitán}, \citenamefont {Gross}, \citenamefont {Saiz}, \citenamefont {Maire}, \citenamefont {Lehmann}, \citenamefont {Rillig}, \citenamefont {Solé},\ and\ \citenamefont {Maestre}}]{Berdugo_2020}%
  \BibitemOpen
  \bibfield  {author} {\bibinfo {author} {\bibfnamefont {M.}~\bibnamefont {Berdugo}}, \bibinfo {author} {\bibfnamefont {M.}~\bibnamefont {Delgado-Baquerizo}}, \bibinfo {author} {\bibfnamefont {S.}~\bibnamefont {Soliveres}}, \bibinfo {author} {\bibfnamefont {R.}~\bibnamefont {Hernández-Clemente}}, \bibinfo {author} {\bibfnamefont {Y.}~\bibnamefont {Zhao}}, \bibinfo {author} {\bibfnamefont {J.~J.}\ \bibnamefont {Gaitán}}, \bibinfo {author} {\bibfnamefont {N.}~\bibnamefont {Gross}}, \bibinfo {author} {\bibfnamefont {H.}~\bibnamefont {Saiz}}, \bibinfo {author} {\bibfnamefont {V.}~\bibnamefont {Maire}}, \bibinfo {author} {\bibfnamefont {A.}~\bibnamefont {Lehmann}}, \bibinfo {author} {\bibfnamefont {M.~C.}\ \bibnamefont {Rillig}}, \bibinfo {author} {\bibfnamefont {R.~V.}\ \bibnamefont {Solé}},\ and\ \bibinfo {author} {\bibfnamefont {F.~T.}\ \bibnamefont {Maestre}},\ }\bibfield  {title} {\bibinfo {title} {Global ecosystem thresholds driven by aridity},\ }\href {https://doi.org/10.1126/science.aay5958} {\bibfield
  {journal} {\bibinfo  {journal} {Science}\ }\textbf {\bibinfo {volume} {367}},\ \bibinfo {pages} {787} (\bibinfo {year} {2020})}\BibitemShut {NoStop}%
\bibitem [{\citenamefont {Flores}\ \emph {et~al.}(2024)\citenamefont {Flores}, \citenamefont {Montoya}, \citenamefont {Sakschewski}, \citenamefont {Nascimento}, \citenamefont {Staal}, \citenamefont {Betts}, \citenamefont {Levis}, \citenamefont {Lapola}, \citenamefont {Esquível-Muelbert}, \citenamefont {Jakovac}, \citenamefont {Nobre}, \citenamefont {Oliveira}, \citenamefont {Borma}, \citenamefont {Nian}, \citenamefont {Boers}, \citenamefont {Hecht}, \citenamefont {ter Steege}, \citenamefont {Arieira}, \citenamefont {Lucas}, \citenamefont {Berenguer}, \citenamefont {Marengo}, \citenamefont {Gatti}, \citenamefont {Mattos},\ and\ \citenamefont {Hirota}}]{Flores_2024}%
  \BibitemOpen
  \bibfield  {author} {\bibinfo {author} {\bibfnamefont {B.~M.}\ \bibnamefont {Flores}}, \bibinfo {author} {\bibfnamefont {E.}~\bibnamefont {Montoya}}, \bibinfo {author} {\bibfnamefont {B.}~\bibnamefont {Sakschewski}}, \bibinfo {author} {\bibfnamefont {N.}~\bibnamefont {Nascimento}}, \bibinfo {author} {\bibfnamefont {A.}~\bibnamefont {Staal}}, \bibinfo {author} {\bibfnamefont {R.~A.}\ \bibnamefont {Betts}}, \bibinfo {author} {\bibfnamefont {C.}~\bibnamefont {Levis}}, \bibinfo {author} {\bibfnamefont {D.~M.}\ \bibnamefont {Lapola}}, \bibinfo {author} {\bibfnamefont {A.}~\bibnamefont {Esquível-Muelbert}}, \bibinfo {author} {\bibfnamefont {C.}~\bibnamefont {Jakovac}}, \bibinfo {author} {\bibfnamefont {C.~A.}\ \bibnamefont {Nobre}}, \bibinfo {author} {\bibfnamefont {R.~S.}\ \bibnamefont {Oliveira}}, \bibinfo {author} {\bibfnamefont {L.~S.}\ \bibnamefont {Borma}}, \bibinfo {author} {\bibfnamefont {D.}~\bibnamefont {Nian}}, \bibinfo {author} {\bibfnamefont {N.}~\bibnamefont {Boers}}, \bibinfo {author}
  {\bibfnamefont {S.~B.}\ \bibnamefont {Hecht}}, \bibinfo {author} {\bibfnamefont {H.}~\bibnamefont {ter Steege}}, \bibinfo {author} {\bibfnamefont {J.}~\bibnamefont {Arieira}}, \bibinfo {author} {\bibfnamefont {I.~L.}\ \bibnamefont {Lucas}}, \bibinfo {author} {\bibfnamefont {E.}~\bibnamefont {Berenguer}}, \bibinfo {author} {\bibfnamefont {J.~A.}\ \bibnamefont {Marengo}}, \bibinfo {author} {\bibfnamefont {L.~V.}\ \bibnamefont {Gatti}}, \bibinfo {author} {\bibfnamefont {C.~R.~C.}\ \bibnamefont {Mattos}},\ and\ \bibinfo {author} {\bibfnamefont {M.}~\bibnamefont {Hirota}},\ }\bibfield  {title} {\bibinfo {title} {Critical transitions in the amazon forest system},\ }\href {https://doi.org/10.1038/s41586-023-06970-0} {\bibfield  {journal} {\bibinfo  {journal} {Nature}\ }\textbf {\bibinfo {volume} {626}},\ \bibinfo {pages} {555} (\bibinfo {year} {2024})}\BibitemShut {NoStop}%
\bibitem [{\citenamefont {Yang}\ \emph {et~al.}(2025)\citenamefont {Yang}, \citenamefont {Liang},\ and\ \citenamefont {Zhou}}]{Yang_2025}%
  \BibitemOpen
  \bibfield  {author} {\bibinfo {author} {\bibfnamefont {Z.}~\bibnamefont {Yang}}, \bibinfo {author} {\bibfnamefont {J.}~\bibnamefont {Liang}},\ and\ \bibinfo {author} {\bibfnamefont {C.}~\bibnamefont {Zhou}},\ }\bibfield  {title} {\bibinfo {title} {Critical avalanches in excitation-inhibition balanced networks reconcile response reliability with sensitivity for optimal neural representation},\ }\href {https://doi.org/10.1103/PhysRevLett.134.028401} {\bibfield  {journal} {\bibinfo  {journal} {Physical Review Letters}\ }\textbf {\bibinfo {volume} {134}},\ \bibinfo {pages} {028401} (\bibinfo {year} {2025})}\BibitemShut {NoStop}%
\bibitem [{\citenamefont {Moran}\ \emph {et~al.}(2025)\citenamefont {Moran}, \citenamefont {Pijpers}, \citenamefont {Weitzel}, \citenamefont {Bouchaud},\ and\ \citenamefont {Panja}}]{Moran_2025}%
  \BibitemOpen
  \bibfield  {author} {\bibinfo {author} {\bibfnamefont {J.}~\bibnamefont {Moran}}, \bibinfo {author} {\bibfnamefont {F.~P.}\ \bibnamefont {Pijpers}}, \bibinfo {author} {\bibfnamefont {U.}~\bibnamefont {Weitzel}}, \bibinfo {author} {\bibfnamefont {J.-P.}\ \bibnamefont {Bouchaud}},\ and\ \bibinfo {author} {\bibfnamefont {D.}~\bibnamefont {Panja}},\ }\bibfield  {title} {\bibinfo {title} {Critical fragility in sociotechnical systems},\ }\href {https://doi.org/10.1073/pnas.2415139122} {\bibfield  {journal} {\bibinfo  {journal} {Proceedings of the National Academy of Sciences}\ }\textbf {\bibinfo {volume} {122}},\ \bibinfo {pages} {e2415139122} (\bibinfo {year} {2025})}\BibitemShut {NoStop}%
\bibitem [{\citenamefont {Sun}(2000)}]{Sun_2000}%
  \BibitemOpen
  \bibfield  {author} {\bibinfo {author} {\bibfnamefont {J.~Z.}\ \bibnamefont {Sun}},\ }\bibfield  {title} {\bibinfo {title} {Spin-current interaction with a monodomain magnetic body: A model study},\ }\href {https://doi.org/10.1103/PhysRevB.62.570} {\bibfield  {journal} {\bibinfo  {journal} {Physical Review B}\ }\textbf {\bibinfo {volume} {62}},\ \bibinfo {pages} {570} (\bibinfo {year} {2000})}\BibitemShut {NoStop}%
\bibitem [{\citenamefont {Lee}\ \emph {et~al.}(2013)\citenamefont {Lee}, \citenamefont {Lee}, \citenamefont {Min},\ and\ \citenamefont {Lee}}]{Lee_2013}%
  \BibitemOpen
  \bibfield  {author} {\bibinfo {author} {\bibfnamefont {K.-S.}\ \bibnamefont {Lee}}, \bibinfo {author} {\bibfnamefont {S.-W.}\ \bibnamefont {Lee}}, \bibinfo {author} {\bibfnamefont {B.-C.}\ \bibnamefont {Min}},\ and\ \bibinfo {author} {\bibfnamefont {K.-J.}\ \bibnamefont {Lee}},\ }\bibfield  {title} {\bibinfo {title} {Threshold current for switching of a perpendicular magnetic layer induced by spin hall effect},\ }\href {https://doi.org/10.1063/1.4798288} {\bibfield  {journal} {\bibinfo  {journal} {Applied Physics Letters}\ }\textbf {\bibinfo {volume} {102}},\ \bibinfo {pages} {112410} (\bibinfo {year} {2013})}\BibitemShut {NoStop}%
\bibitem [{\citenamefont {Taniguchi}\ \emph {et~al.}(2015)\citenamefont {Taniguchi}, \citenamefont {Mitani},\ and\ \citenamefont {Hayashi}}]{Taniguchi_2015}%
  \BibitemOpen
  \bibfield  {author} {\bibinfo {author} {\bibfnamefont {T.}~\bibnamefont {Taniguchi}}, \bibinfo {author} {\bibfnamefont {S.}~\bibnamefont {Mitani}},\ and\ \bibinfo {author} {\bibfnamefont {M.}~\bibnamefont {Hayashi}},\ }\bibfield  {title} {\bibinfo {title} {Critical current destabilizing perpendicular magnetization by the spin hall effect},\ }\href {https://doi.org/10.1103/PhysRevB.92.024428} {\bibfield  {journal} {\bibinfo  {journal} {Physical Review B}\ }\textbf {\bibinfo {volume} {92}},\ \bibinfo {pages} {024428} (\bibinfo {year} {2015})}\BibitemShut {NoStop}%
\bibitem [{\citenamefont {Taniguchi}\ and\ \citenamefont {Kubota}(2016)}]{Taniguchi_2016}%
  \BibitemOpen
  \bibfield  {author} {\bibinfo {author} {\bibfnamefont {T.}~\bibnamefont {Taniguchi}}\ and\ \bibinfo {author} {\bibfnamefont {H.}~\bibnamefont {Kubota}},\ }\bibfield  {title} {\bibinfo {title} {Instability analysis of spin-torque oscillator with an in-plane magnetized free layer and a perpendicularly magnetized pinned layer},\ }\href {https://doi.org/10.1103/PhysRevB.93.174401} {\bibfield  {journal} {\bibinfo  {journal} {Physical Review B}\ }\textbf {\bibinfo {volume} {93}},\ \bibinfo {pages} {174401} (\bibinfo {year} {2016})}\BibitemShut {NoStop}%
\bibitem [{\citenamefont {Zhu}\ and\ \citenamefont {Zhao}(2020)}]{Zhu_2020}%
  \BibitemOpen
  \bibfield  {author} {\bibinfo {author} {\bibfnamefont {D.}~\bibnamefont {Zhu}}\ and\ \bibinfo {author} {\bibfnamefont {W.}~\bibnamefont {Zhao}},\ }\bibfield  {title} {\bibinfo {title} {Threshold current density for perpendicular magnetization switching through spin-orbit torque},\ }\href {https://doi.org/10.1103/PhysRevApplied.13.044078} {\bibfield  {journal} {\bibinfo  {journal} {Physical Review Applied}\ }\textbf {\bibinfo {volume} {13}},\ \bibinfo {pages} {044078} (\bibinfo {year} {2020})}\BibitemShut {NoStop}%
\bibitem [{\citenamefont {Zhang}\ \emph {et~al.}(2024{\natexlab{a}})\citenamefont {Zhang}, \citenamefont {Xu},\ and\ \citenamefont {Zhu}}]{Zhang_2024_prb}%
  \BibitemOpen
  \bibfield  {author} {\bibinfo {author} {\bibfnamefont {X.}~\bibnamefont {Zhang}}, \bibinfo {author} {\bibfnamefont {Z.}~\bibnamefont {Xu}},\ and\ \bibinfo {author} {\bibfnamefont {Z.}~\bibnamefont {Zhu}},\ }\bibfield  {title} {\bibinfo {title} {Revisiting the analytical solution of spin-orbit torque switched nanoscale perpendicular ferromagnets},\ }\href {https://doi.org/10.1103/PhysRevB.110.184428} {\bibfield  {journal} {\bibinfo  {journal} {Physical Review B}\ }\textbf {\bibinfo {volume} {110}},\ \bibinfo {pages} {184428} (\bibinfo {year} {2024}{\natexlab{a}})}\BibitemShut {NoStop}%
\bibitem [{\citenamefont {Negele}(1982)}]{Negele_1982}%
  \BibitemOpen
  \bibfield  {author} {\bibinfo {author} {\bibfnamefont {J.~W.}\ \bibnamefont {Negele}},\ }\bibfield  {title} {\bibinfo {title} {The mean-field theory of nuclear structure and dynamics},\ }\href {https://doi.org/10.1103/RevModPhys.54.913} {\bibfield  {journal} {\bibinfo  {journal} {Reviews of Modern Physics}\ }\textbf {\bibinfo {volume} {54}},\ \bibinfo {pages} {913} (\bibinfo {year} {1982})}\BibitemShut {NoStop}%
\bibitem [{\citenamefont {Aranson}\ and\ \citenamefont {Kramer}(2002)}]{Aranson_2002}%
  \BibitemOpen
  \bibfield  {author} {\bibinfo {author} {\bibfnamefont {I.~S.}\ \bibnamefont {Aranson}}\ and\ \bibinfo {author} {\bibfnamefont {L.}~\bibnamefont {Kramer}},\ }\bibfield  {title} {\bibinfo {title} {The world of the complex ginzburg-landau equation},\ }\href {https://doi.org/10.1103/RevModPhys.74.99} {\bibfield  {journal} {\bibinfo  {journal} {Reviews of Modern Physics}\ }\textbf {\bibinfo {volume} {74}},\ \bibinfo {pages} {99} (\bibinfo {year} {2002})}\BibitemShut {NoStop}%
\bibitem [{\citenamefont {Nienhuis}\ and\ \citenamefont {Nauenberg}(1975)}]{Nienhuis_1975}%
  \BibitemOpen
  \bibfield  {author} {\bibinfo {author} {\bibfnamefont {B.}~\bibnamefont {Nienhuis}}\ and\ \bibinfo {author} {\bibfnamefont {M.}~\bibnamefont {Nauenberg}},\ }\bibfield  {title} {\bibinfo {title} {First-order phase transitions in renormalization-group theory},\ }\href {https://doi.org/10.1103/PhysRevLett.35.477} {\bibfield  {journal} {\bibinfo  {journal} {Physical Review Letters}\ }\textbf {\bibinfo {volume} {35}},\ \bibinfo {pages} {477} (\bibinfo {year} {1975})}\BibitemShut {NoStop}%
\bibitem [{\citenamefont {Wang}(2016)}]{Wang_2016}%
  \BibitemOpen
  \bibfield  {author} {\bibinfo {author} {\bibfnamefont {L.}~\bibnamefont {Wang}},\ }\bibfield  {title} {\bibinfo {title} {Discovering phase transitions with unsupervised learning},\ }\href {https://doi.org/10.1103/PhysRevB.94.195105} {\bibfield  {journal} {\bibinfo  {journal} {Physical Review B}\ }\textbf {\bibinfo {volume} {94}},\ \bibinfo {pages} {195105} (\bibinfo {year} {2016})}\BibitemShut {NoStop}%
\bibitem [{\citenamefont {Huang}\ \emph {et~al.}(2024)\citenamefont {Huang}, \citenamefont {Bathiany}, \citenamefont {Ashwin},\ and\ \citenamefont {Boers}}]{Huang_2024}%
  \BibitemOpen
  \bibfield  {author} {\bibinfo {author} {\bibfnamefont {Y.}~\bibnamefont {Huang}}, \bibinfo {author} {\bibfnamefont {S.}~\bibnamefont {Bathiany}}, \bibinfo {author} {\bibfnamefont {P.}~\bibnamefont {Ashwin}},\ and\ \bibinfo {author} {\bibfnamefont {N.}~\bibnamefont {Boers}},\ }\bibfield  {title} {\bibinfo {title} {Deep learning for predicting rate-induced tipping},\ }\href {https://doi.org/10.1038/s42256-024-00937-0} {\bibfield  {journal} {\bibinfo  {journal} {Nature Machine Intelligence}\ }\textbf {\bibinfo {volume} {6}},\ \bibinfo {pages} {1556} (\bibinfo {year} {2024})}\BibitemShut {NoStop}%
\bibitem [{\citenamefont {Panahi}\ \emph {et~al.}(2024)\citenamefont {Panahi}, \citenamefont {Kong}, \citenamefont {Moradi}, \citenamefont {Zhai}, \citenamefont {Glaz}, \citenamefont {Haile},\ and\ \citenamefont {Lai}}]{Panahi_2024}%
  \BibitemOpen
  \bibfield  {author} {\bibinfo {author} {\bibfnamefont {S.}~\bibnamefont {Panahi}}, \bibinfo {author} {\bibfnamefont {L.-W.}\ \bibnamefont {Kong}}, \bibinfo {author} {\bibfnamefont {M.}~\bibnamefont {Moradi}}, \bibinfo {author} {\bibfnamefont {Z.-M.}\ \bibnamefont {Zhai}}, \bibinfo {author} {\bibfnamefont {B.}~\bibnamefont {Glaz}}, \bibinfo {author} {\bibfnamefont {M.}~\bibnamefont {Haile}},\ and\ \bibinfo {author} {\bibfnamefont {Y.-C.}\ \bibnamefont {Lai}},\ }\bibfield  {title} {\bibinfo {title} {Machine learning prediction of tipping in complex dynamical systems},\ }\href {https://doi.org/10.1103/PhysRevResearch.6.043194} {\bibfield  {journal} {\bibinfo  {journal} {Physical Review Research}\ }\textbf {\bibinfo {volume} {6}},\ \bibinfo {pages} {043194} (\bibinfo {year} {2024})}\BibitemShut {NoStop}%
\bibitem [{\citenamefont {Scheffer}\ \emph {et~al.}(2009)\citenamefont {Scheffer}, \citenamefont {Bascompte}, \citenamefont {Brock}, \citenamefont {Brovkin}, \citenamefont {Carpenter}, \citenamefont {Dakos}, \citenamefont {Held}, \citenamefont {van Nes}, \citenamefont {Rietkerk},\ and\ \citenamefont {Sugihara}}]{Scheffer_2009}%
  \BibitemOpen
  \bibfield  {author} {\bibinfo {author} {\bibfnamefont {M.}~\bibnamefont {Scheffer}}, \bibinfo {author} {\bibfnamefont {J.}~\bibnamefont {Bascompte}}, \bibinfo {author} {\bibfnamefont {W.~A.}\ \bibnamefont {Brock}}, \bibinfo {author} {\bibfnamefont {V.}~\bibnamefont {Brovkin}}, \bibinfo {author} {\bibfnamefont {S.~R.}\ \bibnamefont {Carpenter}}, \bibinfo {author} {\bibfnamefont {V.}~\bibnamefont {Dakos}}, \bibinfo {author} {\bibfnamefont {H.}~\bibnamefont {Held}}, \bibinfo {author} {\bibfnamefont {E.~H.}\ \bibnamefont {van Nes}}, \bibinfo {author} {\bibfnamefont {M.}~\bibnamefont {Rietkerk}},\ and\ \bibinfo {author} {\bibfnamefont {G.}~\bibnamefont {Sugihara}},\ }\bibfield  {title} {\bibinfo {title} {Early-warning signals for critical transitions},\ }\href {https://doi.org/10.1038/nature08227} {\bibfield  {journal} {\bibinfo  {journal} {Nature}\ }\textbf {\bibinfo {volume} {461}},\ \bibinfo {pages} {53} (\bibinfo {year} {2009})}\BibitemShut {NoStop}%
\bibitem [{\citenamefont {Scheffer}\ \emph {et~al.}(2012)\citenamefont {Scheffer}, \citenamefont {Carpenter}, \citenamefont {Lenton}, \citenamefont {Bascompte}, \citenamefont {Brock}, \citenamefont {Dakos}, \citenamefont {van~de Koppel}, \citenamefont {van~de Leemput}, \citenamefont {Levin}, \citenamefont {van Nes}, \citenamefont {Pascual},\ and\ \citenamefont {Vandermeer}}]{Scheffer_2012}%
  \BibitemOpen
  \bibfield  {author} {\bibinfo {author} {\bibfnamefont {M.}~\bibnamefont {Scheffer}}, \bibinfo {author} {\bibfnamefont {S.~R.}\ \bibnamefont {Carpenter}}, \bibinfo {author} {\bibfnamefont {T.~M.}\ \bibnamefont {Lenton}}, \bibinfo {author} {\bibfnamefont {J.}~\bibnamefont {Bascompte}}, \bibinfo {author} {\bibfnamefont {W.}~\bibnamefont {Brock}}, \bibinfo {author} {\bibfnamefont {V.}~\bibnamefont {Dakos}}, \bibinfo {author} {\bibfnamefont {J.}~\bibnamefont {van~de Koppel}}, \bibinfo {author} {\bibfnamefont {I.~A.}\ \bibnamefont {van~de Leemput}}, \bibinfo {author} {\bibfnamefont {S.~A.}\ \bibnamefont {Levin}}, \bibinfo {author} {\bibfnamefont {E.~H.}\ \bibnamefont {van Nes}}, \bibinfo {author} {\bibfnamefont {M.}~\bibnamefont {Pascual}},\ and\ \bibinfo {author} {\bibfnamefont {J.}~\bibnamefont {Vandermeer}},\ }\bibfield  {title} {\bibinfo {title} {Anticipating critical transitions},\ }\href {https://doi.org/10.1126/science.1225244} {\bibfield  {journal} {\bibinfo  {journal} {Science}\ }\textbf {\bibinfo
  {volume} {338}},\ \bibinfo {pages} {344} (\bibinfo {year} {2012})}\BibitemShut {NoStop}%
\bibitem [{\citenamefont {Ralph}\ and\ \citenamefont {Stiles}(2008)}]{Ralph_2008}%
  \BibitemOpen
  \bibfield  {author} {\bibinfo {author} {\bibfnamefont {D.~C.}\ \bibnamefont {Ralph}}\ and\ \bibinfo {author} {\bibfnamefont {M.~D.}\ \bibnamefont {Stiles}},\ }\bibfield  {title} {\bibinfo {title} {Spin transfer torques},\ }\href {https://doi.org/https://doi.org/10.1016/j.jmmm.2007.12.019} {\bibfield  {journal} {\bibinfo  {journal} {Journal of Magnetism and Magnetic Materials}\ }\textbf {\bibinfo {volume} {320}},\ \bibinfo {pages} {1190} (\bibinfo {year} {2008})}\BibitemShut {NoStop}%
\bibitem [{\citenamefont {Song}\ \emph {et~al.}(2021)\citenamefont {Song}, \citenamefont {Zhang}, \citenamefont {Liao}, \citenamefont {Zhou}, \citenamefont {Zhou}, \citenamefont {Chen}, \citenamefont {You}, \citenamefont {Chen},\ and\ \citenamefont {Pan}}]{Song_2021}%
  \BibitemOpen
  \bibfield  {author} {\bibinfo {author} {\bibfnamefont {C.}~\bibnamefont {Song}}, \bibinfo {author} {\bibfnamefont {R.}~\bibnamefont {Zhang}}, \bibinfo {author} {\bibfnamefont {L.}~\bibnamefont {Liao}}, \bibinfo {author} {\bibfnamefont {Y.}~\bibnamefont {Zhou}}, \bibinfo {author} {\bibfnamefont {X.}~\bibnamefont {Zhou}}, \bibinfo {author} {\bibfnamefont {R.}~\bibnamefont {Chen}}, \bibinfo {author} {\bibfnamefont {Y.}~\bibnamefont {You}}, \bibinfo {author} {\bibfnamefont {X.}~\bibnamefont {Chen}},\ and\ \bibinfo {author} {\bibfnamefont {F.}~\bibnamefont {Pan}},\ }\bibfield  {title} {\bibinfo {title} {Spin-orbit torques: Materials, mechanisms, performances, and potential applications},\ }\href {https://doi.org/https://doi.org/10.1016/j.pmatsci.2020.100761} {\bibfield  {journal} {\bibinfo  {journal} {Progress in Materials Science}\ }\textbf {\bibinfo {volume} {118}},\ \bibinfo {pages} {100761} (\bibinfo {year} {2021})}\BibitemShut {NoStop}%
\bibitem [{\citenamefont {Han}\ \emph {et~al.}(2021)\citenamefont {Han}, \citenamefont {Wang}, \citenamefont {Wan}, \citenamefont {Yu},\ and\ \citenamefont {Lv}}]{Han_2021}%
  \BibitemOpen
  \bibfield  {author} {\bibinfo {author} {\bibfnamefont {X.}~\bibnamefont {Han}}, \bibinfo {author} {\bibfnamefont {X.}~\bibnamefont {Wang}}, \bibinfo {author} {\bibfnamefont {C.}~\bibnamefont {Wan}}, \bibinfo {author} {\bibfnamefont {G.}~\bibnamefont {Yu}},\ and\ \bibinfo {author} {\bibfnamefont {X.}~\bibnamefont {Lv}},\ }\bibfield  {title} {\bibinfo {title} {Spin-orbit torques: Materials, physics, and devices},\ }\href {https://doi.org/10.1063/5.0039147} {\bibfield  {journal} {\bibinfo  {journal} {Applied Physics Letters}\ }\textbf {\bibinfo {volume} {118}},\ \bibinfo {pages} {120502} (\bibinfo {year} {2021})}\BibitemShut {NoStop}%
\bibitem [{\citenamefont {Kirilyuk}\ \emph {et~al.}(2013)\citenamefont {Kirilyuk}, \citenamefont {Kimel},\ and\ \citenamefont {Rasing}}]{Kirilyuk_2013}%
  \BibitemOpen
  \bibfield  {author} {\bibinfo {author} {\bibfnamefont {A.}~\bibnamefont {Kirilyuk}}, \bibinfo {author} {\bibfnamefont {A.~V.}\ \bibnamefont {Kimel}},\ and\ \bibinfo {author} {\bibfnamefont {T.}~\bibnamefont {Rasing}},\ }\bibfield  {title} {\bibinfo {title} {Laser-induced magnetization dynamics and reversal in ferrimagnetic alloys},\ }\href {https://doi.org/10.1088/0034-4885/76/2/026501} {\bibfield  {journal} {\bibinfo  {journal} {Rep Prog Phys}\ }\textbf {\bibinfo {volume} {76}},\ \bibinfo {pages} {026501} (\bibinfo {year} {2013})}\BibitemShut {NoStop}%
\bibitem [{\citenamefont {Cheng}\ \emph {et~al.}(2017)\citenamefont {Cheng}, \citenamefont {Li}, \citenamefont {Wang}, \citenamefont {Cheng},\ and\ \citenamefont {Miao}}]{Cheng_2017}%
  \BibitemOpen
  \bibfield  {author} {\bibinfo {author} {\bibfnamefont {W.}~\bibnamefont {Cheng}}, \bibinfo {author} {\bibfnamefont {X.}~\bibnamefont {Li}}, \bibinfo {author} {\bibfnamefont {H.}~\bibnamefont {Wang}}, \bibinfo {author} {\bibfnamefont {X.}~\bibnamefont {Cheng}},\ and\ \bibinfo {author} {\bibfnamefont {X.}~\bibnamefont {Miao}},\ }\bibfield  {title} {\bibinfo {title} {Laser induced ultrafast magnetization reversal in tbco film},\ }\href {https://doi.org/10.1063/1.4975659} {\bibfield  {journal} {\bibinfo  {journal} {AIP Advances}\ }\textbf {\bibinfo {volume} {7}},\ \bibinfo {pages} {056018} (\bibinfo {year} {2017})}\BibitemShut {NoStop}%
\bibitem [{\citenamefont {Deb}\ \emph {et~al.}(2018)\citenamefont {Deb}, \citenamefont {Molho}, \citenamefont {Barbara},\ and\ \citenamefont {Bigot}}]{Deb_2018}%
  \BibitemOpen
  \bibfield  {author} {\bibinfo {author} {\bibfnamefont {M.}~\bibnamefont {Deb}}, \bibinfo {author} {\bibfnamefont {P.}~\bibnamefont {Molho}}, \bibinfo {author} {\bibfnamefont {B.}~\bibnamefont {Barbara}},\ and\ \bibinfo {author} {\bibfnamefont {J.-Y.}\ \bibnamefont {Bigot}},\ }\bibfield  {title} {\bibinfo {title} {Controlling laser-induced magnetization reversal dynamics in a rare-earth iron garnet across the magnetization compensation point},\ }\href {https://doi.org/10.1103/PhysRevB.97.134419} {\bibfield  {journal} {\bibinfo  {journal} {Physical Review B}\ }\textbf {\bibinfo {volume} {97}},\ \bibinfo {pages} {134419} (\bibinfo {year} {2018})}\BibitemShut {NoStop}%
\bibitem [{\citenamefont {Peng}\ \emph {et~al.}(2023)\citenamefont {Peng}, \citenamefont {Salomoni}, \citenamefont {Malinowski}, \citenamefont {Zhang}, \citenamefont {Hohlfeld}, \citenamefont {Buda-Prejbeanu}, \citenamefont {Gorchon}, \citenamefont {Vergès}, \citenamefont {Lin}, \citenamefont {Lacour}, \citenamefont {Sousa}, \citenamefont {Prejbeanu}, \citenamefont {Mangin},\ and\ \citenamefont {Hehn}}]{Peng_2023}%
  \BibitemOpen
  \bibfield  {author} {\bibinfo {author} {\bibfnamefont {Y.}~\bibnamefont {Peng}}, \bibinfo {author} {\bibfnamefont {D.}~\bibnamefont {Salomoni}}, \bibinfo {author} {\bibfnamefont {G.}~\bibnamefont {Malinowski}}, \bibinfo {author} {\bibfnamefont {W.}~\bibnamefont {Zhang}}, \bibinfo {author} {\bibfnamefont {J.}~\bibnamefont {Hohlfeld}}, \bibinfo {author} {\bibfnamefont {L.~D.}\ \bibnamefont {Buda-Prejbeanu}}, \bibinfo {author} {\bibfnamefont {J.}~\bibnamefont {Gorchon}}, \bibinfo {author} {\bibfnamefont {M.}~\bibnamefont {Vergès}}, \bibinfo {author} {\bibfnamefont {J.~X.}\ \bibnamefont {Lin}}, \bibinfo {author} {\bibfnamefont {D.}~\bibnamefont {Lacour}}, \bibinfo {author} {\bibfnamefont {R.~C.}\ \bibnamefont {Sousa}}, \bibinfo {author} {\bibfnamefont {I.~L.}\ \bibnamefont {Prejbeanu}}, \bibinfo {author} {\bibfnamefont {S.}~\bibnamefont {Mangin}},\ and\ \bibinfo {author} {\bibfnamefont {M.}~\bibnamefont {Hehn}},\ }\bibfield  {title} {\bibinfo {title} {In-plane reorientation induced single laser pulse
  magnetization reversal},\ }\href {https://doi.org/10.1038/s41467-023-40721-z} {\bibfield  {journal} {\bibinfo  {journal} {Nature Communications}\ }\textbf {\bibinfo {volume} {14}},\ \bibinfo {pages} {5000} (\bibinfo {year} {2023})}\BibitemShut {NoStop}%
\bibitem [{\citenamefont {Lin}\ \emph {et~al.}(2024)\citenamefont {Lin}, \citenamefont {Le~Guen}, \citenamefont {Hohlfeld}, \citenamefont {Igarashi}, \citenamefont {Remy}, \citenamefont {Gorchon}, \citenamefont {Malinowski}, \citenamefont {Mangin}, \citenamefont {Hauet},\ and\ \citenamefont {Hehn}}]{Lin_2024}%
  \BibitemOpen
  \bibfield  {author} {\bibinfo {author} {\bibfnamefont {J.-X.}\ \bibnamefont {Lin}}, \bibinfo {author} {\bibfnamefont {Y.}~\bibnamefont {Le~Guen}}, \bibinfo {author} {\bibfnamefont {J.}~\bibnamefont {Hohlfeld}}, \bibinfo {author} {\bibfnamefont {J.}~\bibnamefont {Igarashi}}, \bibinfo {author} {\bibfnamefont {Q.}~\bibnamefont {Remy}}, \bibinfo {author} {\bibfnamefont {J.}~\bibnamefont {Gorchon}}, \bibinfo {author} {\bibfnamefont {G.}~\bibnamefont {Malinowski}}, \bibinfo {author} {\bibfnamefont {S.}~\bibnamefont {Mangin}}, \bibinfo {author} {\bibfnamefont {T.}~\bibnamefont {Hauet}},\ and\ \bibinfo {author} {\bibfnamefont {M.}~\bibnamefont {Hehn}},\ }\bibfield  {title} {\bibinfo {title} {Femtosecond-laser-induced reversal in in-plane-magnetized spin valves},\ }\href {https://doi.org/10.1103/PhysRevApplied.22.044051} {\bibfield  {journal} {\bibinfo  {journal} {Physical Review Applied}\ }\textbf {\bibinfo {volume} {22}},\ \bibinfo {pages} {044051} (\bibinfo {year} {2024})}\BibitemShut {NoStop}%
\bibitem [{sup()}]{sup2025}%
  \BibitemOpen
  \bibfield  {title} {\bibinfo {title} {See supplemental material at [url] for details in response function in the case of magnetization switched by magnetic field, analytical value of perturbation-induced variance and determine the order of the phase transition and the critical exponent based on the perturbation-induced variance and response function in the case of magnetization switched by spin transfer torque},\ }\href@noop {} {\ }\BibitemShut {NoStop}%
\bibitem [{\citenamefont {Mathews}\ \emph {et~al.}(2021)\citenamefont {Mathews}, \citenamefont {Musi},\ and\ \citenamefont {Charipar}}]{Mathews_2021}%
  \BibitemOpen
  \bibfield  {author} {\bibinfo {author} {\bibfnamefont {S.~A.}\ \bibnamefont {Mathews}}, \bibinfo {author} {\bibfnamefont {C.}~\bibnamefont {Musi}},\ and\ \bibinfo {author} {\bibfnamefont {N.}~\bibnamefont {Charipar}},\ }\bibfield  {title} {\bibinfo {title} {Transverse susceptibility of nickel thin films with uniaxial anisotropy},\ }\href {https://www.nature.com/articles/s41598-021-82949-z} {\bibfield  {journal} {\bibinfo  {journal} {Scientific Reports}\ }\textbf {\bibinfo {volume} {11}},\ \bibinfo {pages} {3155} (\bibinfo {year} {2021})}\BibitemShut {NoStop}%
\bibitem [{\citenamefont {Jungfleisch}\ \emph {et~al.}(2015)\citenamefont {Jungfleisch}, \citenamefont {Chumak}, \citenamefont {Kehlberger}, \citenamefont {Lauer}, \citenamefont {Kim}, \citenamefont {Onbasli}, \citenamefont {Ross}, \citenamefont {Kl{\"a}ui},\ and\ \citenamefont {Hillebrands}}]{Jungfleisch_2015}%
  \BibitemOpen
  \bibfield  {author} {\bibinfo {author} {\bibfnamefont {M.}~\bibnamefont {Jungfleisch}}, \bibinfo {author} {\bibfnamefont {A.}~\bibnamefont {Chumak}}, \bibinfo {author} {\bibfnamefont {A.}~\bibnamefont {Kehlberger}}, \bibinfo {author} {\bibfnamefont {V.}~\bibnamefont {Lauer}}, \bibinfo {author} {\bibfnamefont {D.}~\bibnamefont {Kim}}, \bibinfo {author} {\bibfnamefont {M.}~\bibnamefont {Onbasli}}, \bibinfo {author} {\bibfnamefont {C.}~\bibnamefont {Ross}}, \bibinfo {author} {\bibfnamefont {M.}~\bibnamefont {Kl{\"a}ui}},\ and\ \bibinfo {author} {\bibfnamefont {B.}~\bibnamefont {Hillebrands}},\ }\bibfield  {title} {\bibinfo {title} {Thickness and power dependence of the spin-pumping effect in y 3 fe 5 o 12/pt heterostructures measured by the inverse spin hall effect},\ }\href {https://doi.org/10.1103/PhysRevB.91.134407} {\bibfield  {journal} {\bibinfo  {journal} {Physical Review B}\ }\textbf {\bibinfo {volume} {91}},\ \bibinfo {pages} {134407} (\bibinfo {year} {2015})}\BibitemShut {NoStop}%
\bibitem [{\citenamefont {Chantrell}(1989)}]{Chantrell_1989}%
  \BibitemOpen
  \bibfield  {author} {\bibinfo {author} {\bibfnamefont {R.}~\bibnamefont {Chantrell}},\ }\bibfield  {title} {\bibinfo {title} {The magnetic properties of textured fine particle systems},\ }\href {https://doi.org/10.1155/TSM.11.107} {\bibfield  {journal} {\bibinfo  {journal} {Texture, Stress, and Microstructure}\ }\textbf {\bibinfo {volume} {11}},\ \bibinfo {pages} {107} (\bibinfo {year} {1989})}\BibitemShut {NoStop}%
\bibitem [{\citenamefont {Cookson}(2002)}]{Cookson_2002}%
  \BibitemOpen
  \bibfield  {author} {\bibinfo {author} {\bibfnamefont {R.~D.}\ \bibnamefont {Cookson}},\ }\emph {\bibinfo {title} {Transverse susceptibility studies of recording media}},\ \href {https://clok.uclan.ac.uk/id/eprint/7714/3/Richard%20David%20Cookson%20Dec02%20transverse%20%20susceptibility%20studies%20of%20recording%20media%20Degree%20of%20Doctor%20of%20Philosophy%20unpublished%20Dec02%20University%20of%20Central%20Lancashire%20unknown%20263.pdf} {Ph.D. thesis},\ \bibinfo  {school} {University of Central Lancashire} (\bibinfo {year} {2002})\BibitemShut {NoStop}%
\bibitem [{\citenamefont {Moorfield}(2020)}]{Moorfield_2020}%
  \BibitemOpen
  \bibfield  {author} {\bibinfo {author} {\bibfnamefont {C.~T.}\ \bibnamefont {Moorfield}},\ }\emph {\bibinfo {title} {An investigation of non-linear transverse susceptibility as a basis for improving the measurement of anisotropy in magnetic powdered samples.}},\ \href {https://clok.uclan.ac.uk/id/eprint/34671/2/34671%20Moorfield%2C%20Conor%2C%20MSc%20Thesis.pdf} {Ph.D. thesis},\ \bibinfo  {school} {University of Central Lancashire} (\bibinfo {year} {2020})\BibitemShut {NoStop}%
\bibitem [{\citenamefont {Yang}(1952)}]{Yang_1952}%
  \BibitemOpen
  \bibfield  {author} {\bibinfo {author} {\bibfnamefont {C.~N.}\ \bibnamefont {Yang}},\ }\bibfield  {title} {\bibinfo {title} {The spontaneous magnetization of a two-dimensional ising model},\ }\href {https://doi.org/10.1103/PhysRev.85.808} {\bibfield  {journal} {\bibinfo  {journal} {Physical Review}\ }\textbf {\bibinfo {volume} {85}},\ \bibinfo {pages} {808} (\bibinfo {year} {1952})}\BibitemShut {NoStop}%
\bibitem [{\citenamefont {Wang}\ \emph {et~al.}(2018)\citenamefont {Wang}, \citenamefont {Cai}, \citenamefont {Zhu}, \citenamefont {Wang}, \citenamefont {Kan}, \citenamefont {Zhao}, \citenamefont {Cao}, \citenamefont {Wang}, \citenamefont {Zhang}, \citenamefont {Zhang} \emph {et~al.}}]{Wang_2018}%
  \BibitemOpen
  \bibfield  {author} {\bibinfo {author} {\bibfnamefont {M.}~\bibnamefont {Wang}}, \bibinfo {author} {\bibfnamefont {W.}~\bibnamefont {Cai}}, \bibinfo {author} {\bibfnamefont {D.}~\bibnamefont {Zhu}}, \bibinfo {author} {\bibfnamefont {Z.}~\bibnamefont {Wang}}, \bibinfo {author} {\bibfnamefont {J.}~\bibnamefont {Kan}}, \bibinfo {author} {\bibfnamefont {Z.}~\bibnamefont {Zhao}}, \bibinfo {author} {\bibfnamefont {K.}~\bibnamefont {Cao}}, \bibinfo {author} {\bibfnamefont {Z.}~\bibnamefont {Wang}}, \bibinfo {author} {\bibfnamefont {Y.}~\bibnamefont {Zhang}}, \bibinfo {author} {\bibfnamefont {T.}~\bibnamefont {Zhang}}, \emph {et~al.},\ }\bibfield  {title} {\bibinfo {title} {Field-free switching of a perpendicular magnetic tunnel junction through the interplay of spin--orbit and spin-transfer torques},\ }\href {https://doi.org/10.1038/s41928-018-0160-7} {\bibfield  {journal} {\bibinfo  {journal} {Nature electronics}\ }\textbf {\bibinfo {volume} {1}},\ \bibinfo {pages} {582} (\bibinfo {year} {2018})}\BibitemShut
  {NoStop}%
\bibitem [{\citenamefont {Petit}\ \emph {et~al.}(2007)\citenamefont {Petit}, \citenamefont {Baraduc}, \citenamefont {Thirion}, \citenamefont {Ebels}, \citenamefont {Liu}, \citenamefont {Li}, \citenamefont {Wang},\ and\ \citenamefont {Dieny}}]{Petit_2007}%
  \BibitemOpen
  \bibfield  {author} {\bibinfo {author} {\bibfnamefont {S.}~\bibnamefont {Petit}}, \bibinfo {author} {\bibfnamefont {C.}~\bibnamefont {Baraduc}}, \bibinfo {author} {\bibfnamefont {C.}~\bibnamefont {Thirion}}, \bibinfo {author} {\bibfnamefont {U.}~\bibnamefont {Ebels}}, \bibinfo {author} {\bibfnamefont {Y.}~\bibnamefont {Liu}}, \bibinfo {author} {\bibfnamefont {M.}~\bibnamefont {Li}}, \bibinfo {author} {\bibfnamefont {P.}~\bibnamefont {Wang}},\ and\ \bibinfo {author} {\bibfnamefont {B.}~\bibnamefont {Dieny}},\ }\bibfield  {title} {\bibinfo {title} {Spin-torque influence on the high-frequency magnetization fluctuations in magnetic tunnel junctions},\ }\href {https://doi.org/10.1103/PhysRevLett.98.077203} {\bibfield  {journal} {\bibinfo  {journal} {Physical review letters}\ }\textbf {\bibinfo {volume} {98}},\ \bibinfo {pages} {077203} (\bibinfo {year} {2007})}\BibitemShut {NoStop}%
\bibitem [{\citenamefont {Roberts}(2008)}]{Roberts_2008}%
  \BibitemOpen
  \bibfield  {author} {\bibinfo {author} {\bibfnamefont {D.~C.}\ \bibnamefont {Roberts}},\ }\bibfield  {title} {\bibinfo {title} {Linear reformulation of the kuramoto model of self-synchronizing coupled oscillators},\ }\href {https://doi.org/10.1103/PhysRevE.77.031114} {\bibfield  {journal} {\bibinfo  {journal} {Physical Review E}\ }\textbf {\bibinfo {volume} {77}},\ \bibinfo {pages} {031114} (\bibinfo {year} {2008})}\BibitemShut {NoStop}%
\bibitem [{\citenamefont {García-Gudiño}\ \emph {et~al.}(2017)\citenamefont {García-Gudiño}, \citenamefont {Landa}, \citenamefont {Mendoza-Temis}, \citenamefont {Albarado-Ibañez}, \citenamefont {Toledo-Roy}, \citenamefont {Morales},\ and\ \citenamefont {Frank}}]{Garc_2017}%
  \BibitemOpen
  \bibfield  {author} {\bibinfo {author} {\bibfnamefont {D.}~\bibnamefont {García-Gudiño}}, \bibinfo {author} {\bibfnamefont {E.}~\bibnamefont {Landa}}, \bibinfo {author} {\bibfnamefont {J.}~\bibnamefont {Mendoza-Temis}}, \bibinfo {author} {\bibfnamefont {A.}~\bibnamefont {Albarado-Ibañez}}, \bibinfo {author} {\bibfnamefont {J.~C.}\ \bibnamefont {Toledo-Roy}}, \bibinfo {author} {\bibfnamefont {I.~O.}\ \bibnamefont {Morales}},\ and\ \bibinfo {author} {\bibfnamefont {A.~J. P.~O.}\ \bibnamefont {Frank}},\ }\bibfield  {title} {\bibinfo {title} {Enhancement of early warning properties in the kuramoto model and in an atrial fibrillation model due to an external perturbation of the system},\ }\href {https://doi.org/10.1371/journal.pone.0181953} {\bibfield  {journal} {\bibinfo  {journal} {PLoS ONE}\ }\textbf {\bibinfo {volume} {12}},\ \bibinfo {pages} {e0181953} (\bibinfo {year} {2017})}\BibitemShut {NoStop}%
\bibitem [{\citenamefont {Zhang}\ \emph {et~al.}(2024{\natexlab{b}})\citenamefont {Zhang}, \citenamefont {Zhang}, \citenamefont {Liu}, \citenamefont {Zhang}, \citenamefont {Shao}, \citenamefont {Li}, \citenamefont {Chen}, \citenamefont {Liu}, \citenamefont {Ma}, \citenamefont {Han}, \citenamefont {Wang}, \citenamefont {Adams}, \citenamefont {Shi},\ and\ \citenamefont {Ding}}]{Zhang_2024}%
  \BibitemOpen
  \bibfield  {author} {\bibinfo {author} {\bibfnamefont {J.}~\bibnamefont {Zhang}}, \bibinfo {author} {\bibfnamefont {L.-H.}\ \bibnamefont {Zhang}}, \bibinfo {author} {\bibfnamefont {B.}~\bibnamefont {Liu}}, \bibinfo {author} {\bibfnamefont {Z.-Y.}\ \bibnamefont {Zhang}}, \bibinfo {author} {\bibfnamefont {S.-Y.}\ \bibnamefont {Shao}}, \bibinfo {author} {\bibfnamefont {Q.}~\bibnamefont {Li}}, \bibinfo {author} {\bibfnamefont {H.-C.}\ \bibnamefont {Chen}}, \bibinfo {author} {\bibfnamefont {Z.-K.}\ \bibnamefont {Liu}}, \bibinfo {author} {\bibfnamefont {Y.}~\bibnamefont {Ma}}, \bibinfo {author} {\bibfnamefont {T.-Y.}\ \bibnamefont {Han}}, \bibinfo {author} {\bibfnamefont {Q.-F.}\ \bibnamefont {Wang}}, \bibinfo {author} {\bibfnamefont {C.~S.}\ \bibnamefont {Adams}}, \bibinfo {author} {\bibfnamefont {B.-S.}\ \bibnamefont {Shi}},\ and\ \bibinfo {author} {\bibfnamefont {D.-S.}\ \bibnamefont {Ding}},\ }\bibfield  {title} {\bibinfo {title} {Early warning signals of the tipping point in strongly interacting rydberg
  atoms},\ }\href {https://doi.org/10.1103/PhysRevLett.133.243601} {\bibfield  {journal} {\bibinfo  {journal} {Physical Review Letters}\ }\textbf {\bibinfo {volume} {133}},\ \bibinfo {pages} {243601} (\bibinfo {year} {2024}{\natexlab{b}})}\BibitemShut {NoStop}%
\end{thebibliography}
\end{document}